\documentclass[sigconf,authordraft]{acmart}
\AtBeginDocument{%
  \providecommand\BibTeX{{%
    \normalfont B\kern-0.5em{\scshape i\kern-0.25em b}\kern-0.8em\TeX}}}

\copyrightyear{2021} 
\acmYear{2021} 
\setcopyright{acmlicensed}\acmConference[ESEM '21]{ACM / IEEE International Symposium on Empirical Software Engineering and Measurement (ESEM)}{October 11--15, 2021}{Bari, Italy}
\acmBooktitle{ACM / IEEE International Symposium on Empirical Software Engineering and Measurement (ESEM) (ESEM '21), October 11--15, 2021, Bari, Italy}
\acmPrice{15.00}
\acmDOI{10.1145/3475716.3475785}
\acmISBN{978-1-4503-8665-4/21/10}

\usepackage{multirow}

\begin{document}

\title{An Empirical Study on Refactoring-Inducing Pull Requests}


\author{Fl{\'a}via Coelho}
\affiliation{%
  \institution{\normalsize Federal University of Campina Grande}
  \streetaddress{}
  \city{Campina Grande}
  \country{Brazil}}
\email{flavia@copin.ufcg.edu.br}

\author{Nikolaos Tsantalis}
\affiliation{%
  \institution{\normalsize Concordia University}
  \streetaddress{}
  \city{Montreal}
  \country{Canada}}
\email{nikolaos.tsantalis@concordia.ca}

\author{Tiago Massoni}
\affiliation{%
  \institution{\normalsize Federal University of Campina Grande}
  \streetaddress{}
  \city{Campina Grande}
  \country{Brazil}}
\email{massoni@computacao.ufcg.edu.br}

\author{Everton L. G. Alves}
\affiliation{%
  \institution{\normalsize Federal University of Campina Grande}
  \streetaddress{}
  \city{Campina Grande}
  \country{Brazil}}
\email{everton@computacao.ufcg.edu.br}

\renewcommand{\shortauthors}{Coelho, Tsantalis, Massoni, and Alves}

\begin{abstract}
    \textbf{Background:} Pull-based development has shaped the practice of Modern Code Review (MCR), in which reviewers can contribute code improvements, such as refactorings, through comments and commits in Pull Requests (PRs).
    Past MCR studies uniformly treat all PRs, regardless of whether they induce refactoring or not.
    We define a PR as \textit{refactoring-inducing}, when refactoring edits are performed after the initial commit(s), as either a result of discussion among reviewers or spontaneous actions carried out by the PR developer.
    \textbf{Aims:} This mixed study (quantitative and qualitative) explores code reviewing-related aspects intending to characterize refactoring-inducing PRs.
    \textbf{Method:} We hypothesize that refactoring-inducing PRs have distinct characteristics than non-refactoring-inducing ones and thus deserve special attention and treatment from researchers, practitioners, and tool builders. To investigate our hypothesis, we mined a sample of 1,845 Apache's merged PRs from GitHub, mined refactoring edits in these PRs, and ran a comparative study between refactoring-inducing and non-refactoring-inducing PRs. We also manually examined 2,096 review comments and 1,891 detected refactorings from 228 refactoring-inducing PRs.
    \textbf{Results:} We found 30.2\% of refactoring-inducing PRs in our sample and that they significantly differ from non-refactoring-inducing ones in terms of number of commits, code churn, number of file changes, number of review comments, length of discussion, and time to merge. However, we found no statistical evidence that the number of reviewers is related to refactoring-inducement.
    Our qualitative analysis revealed that at least one refactoring edit was induced by review in 133 (58.3\%) of the refactoring-inducing PRs examined.
    \textbf{Conclusions:} Our findings suggest directions for researchers, practitioners, and tool builders to improve practices around pull-based code review.
\end{abstract}

\ccsdesc[300]{Software and its engineering~Programming teams}
\ccsdesc[500]{Software and its engineering~Software evolution}

\keywords{refactoring-inducing pull request, code review mining, empirical study}


\maketitle

\section{Introduction}
In \textit{Modern Code Review} (MCR), developers review code changes in a lightweight, tool-assisted, and asynchronous manner \cite{Bacchelli_ICSE_2013}. 
In this context, regular change-based reviewing, in which code improvements are embraced, became an essential practice in the MCR scenario \cite{Bacchelli_ICSE_2013,Rigby_FSE_2013}.
Code changes may comprise new features, bug fixes, or other maintenance tasks, providing potential opportunities for refactorings \cite{Palomba_ICPC_2017}, which in turn form a significant part of the changes \cite{Barnett_ICSE_2015,Tao_FSE_2012}. 
Empirical evidence suggests a distinction between refactoring-dominant changes and other types. For instance, reviewing bug fixes is more time-consuming than reviewing refactorings, since the latter preserve code behavior \cite{Robbes_ICPC_2007}. 
Given the nature of changes significantly affects code review effectiveness \cite{Ram_FSE_2018}, as it directly influences how reviewers perceive the changes, the provision of suitable resources for assisting code review is essential. 

Characterization studies of MCR have been conducted to investigate technical aspects of reviewing \cite{Rigby_ICSE_2008,Rigby_FSE_2013,Rigby_SEM_2014,Baysal_ESE_2016,Izquierdo_MSR_2016,Bosu_TSE_2017,Sadowski_ICSE_2018}, factors leading to useful code review \cite{Bosu_MSR_2015}, circumstances that contribute to code review quality \cite{Kononenko_ICSE_2016}, and general code review patterns in pull-based development \cite{Li_CST_2017}.
Those studies are relevant because MCR is critical in repository-based software development, especially in \textit{Agile software development}, driven by change and collaboration \cite{Agile}. 

In practice, Git \textit{Pull Requests} (PRs) are relevant to MCR as they promote well-defined and collaborative reviewing. 
Through PRs, the code is subject to a review process in which reviewers may suggest improvements before merging the code to the main branch of a repository \cite{Git_Book_2014}. 
Such improvements may take the form of refactorings, resulting from discussions among the PR author and reviewers on code quality issues, including spontaneous actions of the PR author aiming to refine the originally submitted solution.
We hypothesize that PRs that induce refactoring edits have different characteristics from those that do not, as refactoring may involve design and API changes that require more extensive effort, discussion and knowledge of the project.
It is worth clarifying that this study sheds light on refactorings induced by code review (Section \ref{empirical-study-design}) aiming to provide an initial understanding of how review discussions induce such edits.

\textbf{Motivation}: By distinguishing refactoring-inducing from non-refactoring-inducing PRs, we can potentially advance the understanding of code reviewing at the PR level and assist researchers, practitioners, and tool builders in this context. 
No prior MCR studies made a distinction between refactoring-inducing and non-refactoring-inducing PRs, when analyzing their research questions,
which might have affected their findings or discussions. 
For instance, by also regarding refactoring-inducing PRs, Gousios et al. \cite{Gousios_ICSE_2014} and Kononenko et al. \cite{Kononenko_ICSE_2018} could have found different factors influencing the time to merge a PR; Li et al. \cite{Li_CST_2017} could have included refactoring concerns to the multilevel taxonomy for review comments in the pull-based development model; Pascarella et al. \cite{Pascarella_HCI_2018} could have identified further information to perform a proper code review in presence of refactorings; Paix\~ao et al. \cite{Paixao_MSR_2020} could have complemented the study on the reasons for refactorings during code review when analyzing projects in Gerrit; whereas, Pantiuchina et al. \cite{Pantiuchina_SEM_2020} could have different conclusions on the motivations for refactorings in PRs, since they analyzed PRs in which refactorings were detected even in the initial commit (i.e., these refactorings were not induced from reviewer discussions).
In practice, being unaware of refactoring-inducing PRs' characteristics, practitioners and tool builders might miss opportunities to manage better their resources and to assist developers in PRs, respectively.
Moreover, a refactoring-aware notification system could help in allocating reviewers with more knowledge on the design of the refactored code when a PR becomes refactoring-inducing, as design changes caused by refactoring need to be more extensively discussed and agreed upon.

\vspace{-3mm}
\begin{definition}\label{ripr}
  A PR is refactoring-inducing if refactoring edits are performed in subsequent commits after the initial PR commit(s), as a result of the reviewing process or spontaneous improvements by the PR contributor.
  Let $U = \{u\textsubscript{1}, u\textsubscript{2}, ..., u\textsubscript{w}\}$, a set of repositories in GitHub.  
  Each repository $u\textsubscript{q}$, $1 \leq q \leq w$, has a set of pull requests $P(u\textsubscript{q}) = \{p\textsubscript{1}, p\textsubscript{2}, ..., p\textsubscript{m}\}$ over time.
  Each pull request $p\textsubscript{j}$, $1 \leq j \leq m$, has a set of commits $C(p\textsubscript{j}) = \{c\textsubscript{1}, c\textsubscript{2}, ..., c\textsubscript{n}\}$, in which $I(p\textsubscript{j})$ is the set of initial commits included in the PR when it is created, $I(p\textsubscript{j}) \subseteq C(p\textsubscript{j})$.
  A refactoring-inducing pull request is that in which $\exists$ $c\textsubscript{k} \mid R(c\textsubscript{k}) \neq \varnothing$, where $R(c\textsubscript{k})$ denotes the set of refactorings performed in commit $c\textsubscript{k}$ and $|I(p\textsubscript{j})| < k \leq n$. 
\end{definition}
\vspace{-3mm}

\begin{figure}[h]
  \centering
  \includegraphics[width=\linewidth]{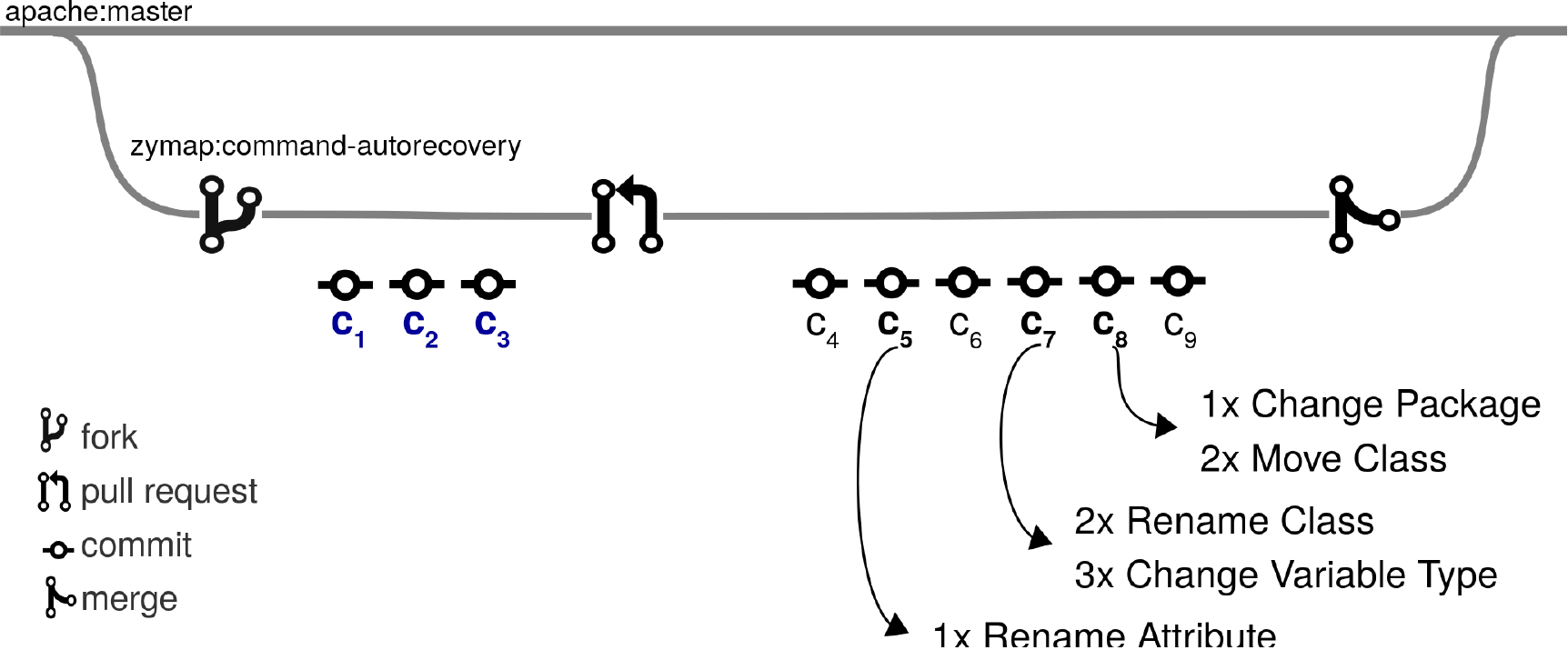}
  \vspace{-7mm}
  \caption{A Refactoring-Inducing Pull Request (Apache bookkeeper PR \#2010), Illustrating Initial Commits ($c_1-c_3$) and Subsequent Commits ($c_4-c_9$).}
  \label{fig:ripr-example}
\end{figure}

To clarify our definition, Figure \ref{fig:ripr-example} depicts a refactoring-inducing PR consisting of three initial commits ($c_1-c_3$) and six subsequent commits ($c_4-c_9$), three of which include refactoring edits ($c_5$, $c_7$, $c_8$), e.g., commit $c_7$ has two \textit{Rename Class} and three \textit{Change Variable Type} refactoring instances. Our study explores differences/similarities between PRs based on the refactorings performed in PR commits subsequent to the initial ones ($c_4-c_9$).

We propose an investigation at the PR level because we understand it as a complete scenario for exploring code reviewing practices in a well-defined scope of development, which allows us to go beyond an investigation at the commit level. 
For instance, we can obtain a global comprehension of contributions to the original code, in terms of both commits and reviewing-related aspects (e.g., reviewers' comments). 
Our conception is mainly inspired by empirical evidence showing that pull-based development is associated with larger numbers of contributions \cite{Zhu_FSE_2016}, and that PR discussions lead to additional refactorings \cite{Pantiuchina_SEM_2020}.
To guide our investigation, we designed the following research questions:

\begin{itemize}[leftmargin=*]
    \item RQ\textsubscript{1}: How common are refactoring-inducing PRs?
    \item RQ\textsubscript{2}: How do refactoring-inducing PRs compare to non-refactoring-inducing ones?
    \item RQ\textsubscript{3}: Are refactoring edits induced by code reviews?
\end{itemize}

\vspace{-2mm}
\vspace{-3.5mm}

We mined merged PRs from Apache's Java repositories in GitHub, and we used state-of-the-art tools and techniques, such as RefactoringMiner \cite{RefactoringMiner} and \textit{Association Rule Learning} (ARL) \cite{Book_MLA_2017} to answer the first two questions.
RefactoringMiner is currently considered the state-of-the-art refactoring detection tool (precision of 97.96\% and recall of 87.2\% \cite{Tsantalis_ICSE_2018}, whereas ARL can discover non-obvious relationships between variables in large datasets \cite{Agrawal_SR_1993}. 
We used RefactoringMiner to detect refactorings in a sample of 1,845 merged PRs.
Then, we performed ARL on two groups (refactoring-inducing and non-refactoring-inducing PRs), and formulated eight (8) hypotheses on differences between refactoring-inducing and non-refactoring-inducing PRs by manually exploring 562 association rules discovered by ARL.
We found that refactoring-inducing PRs significantly differ from non-refactoring-inducing ones in terms of number of subsequent commits, code churn, number of file changes, number of review comments, length of discussion, and time to merge; however, we found no statistical evidence that the number of reviewers is related to refactoring-inducement. 

In order to address the third research question, we carried out a manual investigation of 2,096 review comments cross-referenced to 1,891 detected refactorings from 228 refactoring-inducing PRs -- a stratified sample from our original sample (by considering a confidence level of 95\% and a margin of error of 5\%). We found 133 refactoring-inducing PRs (58.3\%) in which at least one refactoring edit was induced by review comments.

\textbf{Contributions}: 
\begin{enumerate}[leftmargin=*]
\item To the best of our knowledge, this is the first study investigating aspects related to refactoring and code review in the context of refactoring-inducing PRs (Def. \ref{ripr}).
\item We investigate PRs merged by \textit{merge pull request} and \textit{squash and merge} options. We tried to avoid either PRs merged by \textit{rebase and merge} or merged PRs that suffered rebasing, intending to minimize threats to validity (Section \ref{mining_merged_pull_requests}). To deal with squashed commits, we implemented a script that recovers them (\textit{git squash} converts all commits in a PR into a single commit). 
\item We performed a manual analysis of refactoring-inducement, by exploring more than 2,000 review comments.
\item We made available a complete reproduction kit \cite{Reproduction_kit} including the mined dataset and implemented scripts to enable replications and future research.
\end{enumerate}

\section{Background}\label{background}

\subsection{Refactoring and Modern Code Review}
As software evolves to meet new requirements, its code becomes more complex. Throughout this process, design and quality deserve attention \cite{Kataoka_ICSM_2002}. 
For that, code restructurings, coined as refactorings by Opdyke and Johnson \cite{Opdyke_OOPEPA_1990}, are performed to improve the design quality of object-oriented software, while preserving its external behavior, and they should be performed in a structured manner \cite{Fowler_Book_1999,Opdyke_Thesis_1992}. 
Developers can recover those restructurings through refactoring detection tools -- which automatically identify refactoring types applied to the code, for assisting tasks such as studies on code evolution \cite{Palomba_ICPC_2017} and MCR \cite{Alves_TSE_2018,Ge_VLHCC_2017}.
MCR consists of a lightweight code review (in opposition to the formal code inspections specified by Fagan \cite{Fagan_IBM_1976}), tool-assisted, asynchronous, and driven by reviewing code changes, submitted by a developer (author), and manually examined by one or more other developers (reviewers) \cite{Bacchelli_ICSE_2013}. 

\subsection{Git-Based Development and Pull Requests}
\label{merge_types}
Git-based collaborative development as implemented in GitHub \cite{GitHub} has presented a fast growth in the number of developers (more than 56 million) \cite{Octoverse}.
Each Git repository maintains a full history of changes \cite{Git_Book_2014} structured as a linked-list of commits, in turn, organized into multiple lines of development (branches). 
A PR is a commonly used way for submitting contributions to collaboration-based projects \cite{GitHub_PR}. After forking a Git branch, a developer can implement changes, and open a PR to submit them for reviewing in line with the MCR process. Next, reviewers can submit comments based on a diff output that highlights the changes, whereas the author and other contributors can answer the reviewers' comments.
After the reviewing, there are three options of merging:

\vspace{-1mm}
\begin{itemize}[leftmargin=*]
    \item \textit{Merge pull request} merges the PR commits into a \textit{merge commit} and adds them into the main branch, chronologically ordered, as depicted in Figure \ref{fig:commit-merge-type}. Note that the arrows indicate a commit's parent, and the \textit{before} and \textit{after} markers indicate the commits searchable in the PR, respectively, before and after merging;    
    
    \begin{figure}[h]
      \centering
      \includegraphics[width=\linewidth]{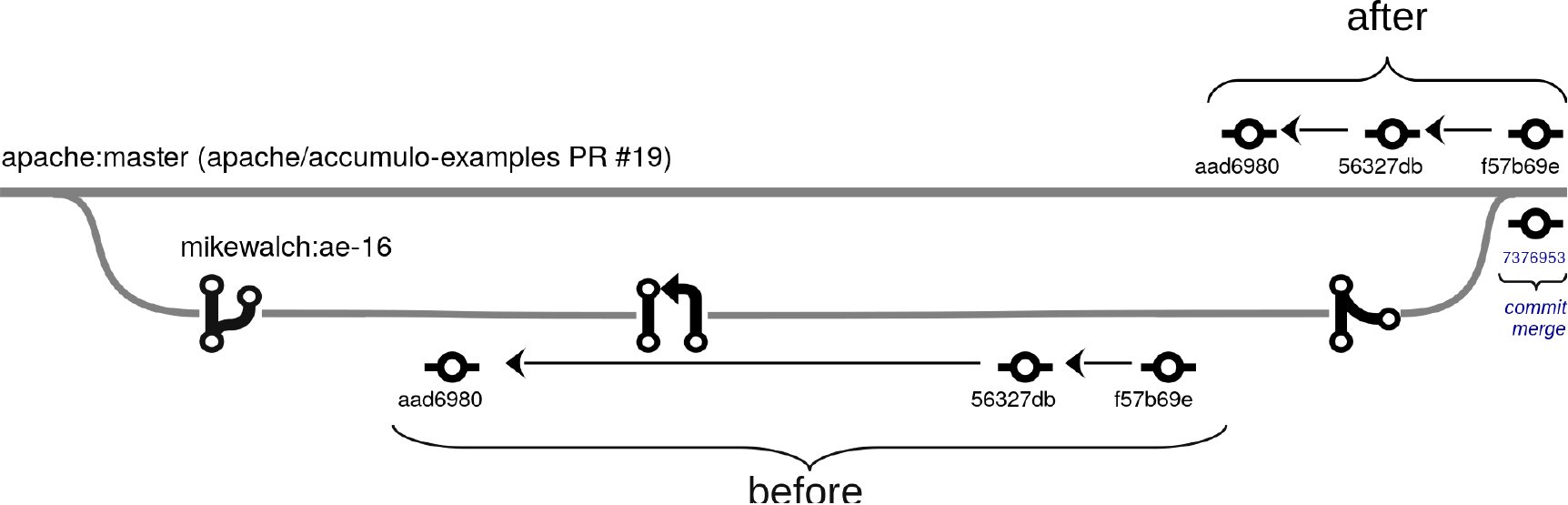}
      \vspace{-7mm}
      \caption{Illustrating Merge Pull Request Option (Apache accumulo-examples PR \#19)}
      \label{fig:commit-merge-type}
    \end{figure}
    
    \item \textit{Squash and merge} squashes the PR commits into a single commit and merges it into the main branch (Figure \ref{fig:squash-merge-type}); and
    
    \item \textit{Rebase and merge} re-writes all commits from one branch onto another, by updating their SHA, in a manner that unwanted history can be discarded, as illustrated in Figure \ref{fig:rebase-type}. In this case, commits \textit{0be3d3f} and \textit{66f02d3} received review comments, but they are not accessible via PR. Hence, it is mandatory to recover the original commits when investigating reviewing-related aspects. Nonetheless, such a recovery is not trivial \cite{Ji_ISSRE_2020}.
    
    \vspace{-2mm}
    \begin{figure}[h]
      \centering
      \includegraphics[width=\linewidth]{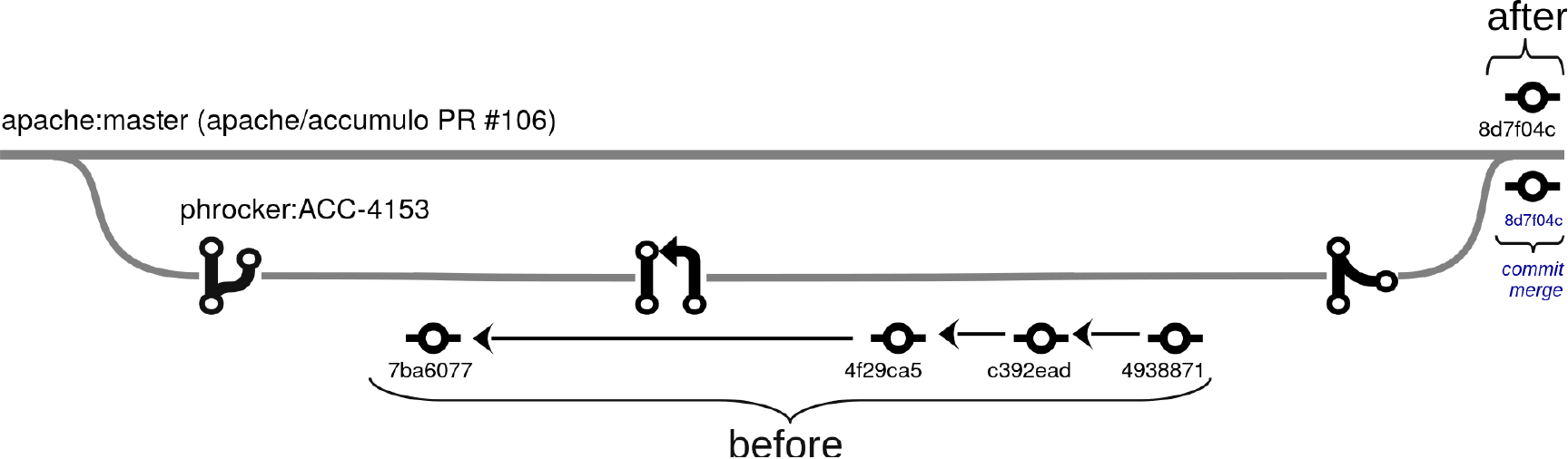}
      \vspace{-7mm}
      \caption{Illustrating Squash and Merge Option (Apache accumulo PR \#106)}
      \label{fig:squash-merge-type}
    \end{figure}
    
    \vspace{-5mm}
    \begin{figure}[h]
      \centering
      \includegraphics[width=\linewidth]{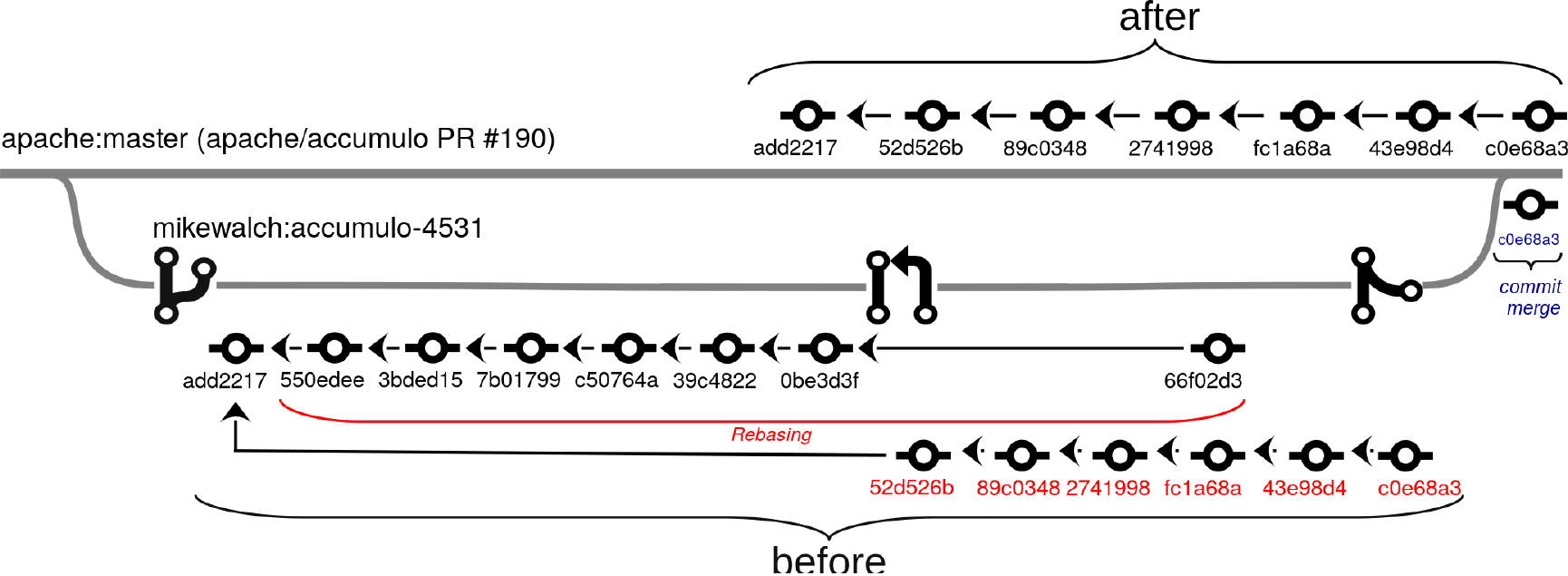}
      \vspace{-7mm}
      \caption{Illustrating Rebase and Merge Option (Apache accumulo PR \#190)}
      \label{fig:rebase-type}
    \end{figure}
    \vspace{-2mm}
\end{itemize}

\subsection{Association Rule Learning}
ARL discovers rules that denote non-obvious relationships between variables in large datasets, e.g., refactoring-inducing PRs with a high number of added lines tend to have a high number of reviewers. 
Formally, let $I = \{i_1, i_2, ..., i_n\}$, a set of n binary attributes (\textit{items}) and $D = \{t_1, t_2, ..., t_m\}$, a set of m transactions (\textit{dataset}), in which each transaction in $D$ consists of items in $I$. Thus, an \textit{Association Rule} (AR) $\{X\} \rightarrow \{Y\}$ indicates the co-occurrence of the tuples $\{X\}$ (\textit{antecedent}) and $\{Y\}$ (\textit{consequent}), where $\{X\},\{Y\} \subseteq I$, $\{X\} \cap \{Y\} = \varnothing$ \cite{Agrawal_SR_1993}.
\textit{Support} 
indicates the number of transactions in $D$ that supports an AR, so expressing its statistical significance.


Interestingness measures can determine the strength of an AR.
\textit{Confidence}
means how likely $\{X\}$ and $\{Y\}$ will occur together. \textit{Lift} 
reveals how X and Y are related to one another (0 denotes no association, $<$ 1 indicates a negative co-occurrence of the antecedent and consequent, and $>$ 1 express that the two occurrences are dependent on one another and the ARs are useful) \cite{Geng_ACS_2006}. 
\textit{Conviction}
is a measure of implication, ranging in the interval $[0, \infty]$. Conviction 1 denotes that antecedent and consequent are unrelated, while $\infty$ expresses logical implications, where confidence is 1 \cite{Brin_CMD_1997}.

ARL usually follows this workflow: feature selection, feature engineering (applying any encoding technique, such as \textit{one-hot encoding} using a group of bits to represent mutually exclusive features \cite{MachineLearning_Book_2018}), algorithm choice and execution, and result interpretation (assisted by interestingness measures) \cite{Xu_DS_2015}.

\section{Motivating Example} \label{motivating-example}
This study has evolved from results of preliminary investigations on refactorings and code reviews to get a better understanding of the topic and plan the research design. 
As a motivating example, we describe a case history, in which we explored the refactoring-inducement and code review aspects.
We randomly selected 24 PRs from Apache's drill repository. 
Then, we ran RefactoringMiner and obtained 11 (45.8\%) refactoring-inducing PRs.

We compared refactoring-inducing and non-refactoring-inducing PRs concerning \textit{code churn} (number of changed lines), and discussion length (i.e., review and non-review comments). As a result, we identified that the refactoring-inducing PRs presented a higher code churn and discussion length than non-refactoring-inducing PRs. Note that we took into account one measure of each context under investigation: changes (code churn), code review (length of discussion), besides the number of refactoring edits.

We manually analysed the refactoring-inducing PRs, by contrasting the descriptions of the detected refactorings by RefactoringMiner against review comments.
Our strategy of analysis consisted of reading comments and searching for keywords (e.g., ``refac'', ``mov'', ``extract'', and ``renam''). 
We observed refactorings directly induced by review comments in four refactoring-inducing PRs. 
To exemplify, in PR \#1762\footnote{Apache drill PR \#1762, available in \url{ https://git.io/JczHh}.}, the review comment ``\textit{Lot of code here and in DefaultMemoryAllocationUtilities are duplicate. May be create a separate MemoryAllocationUtilities to keep the common code...}'' motivated one \textit{Extract Superclass} and four \textit{Pull Up Method} refactorings.

In a nutshell, those results provided insights on the pertinence of (i) exploring technical aspects of changes, code review, and refactorings in the PR level, since we perceived differences between refactoring-inducing and non-refactoring-inducing PRs in terms of code churn and length of discussion; (ii) considering refactorings as part of contributions to the code improvement during code review, and  (iii) investigating quantitatively and qualitatively technical aspects in light of the refactoring-inducing PR definition.

\section{Study Design} \label{empirical-study-design}
The main goal of this study is to investigate code reviewing-related data to characterize refactoring-inducing PRs in Apache's repositories hosted in GitHub, from the reviewers' perspective.
Thus, we formulated these \textbf{research questions}:

\begin{itemize}[leftmargin=*]
    \item RQ\textsubscript{1}: How common are refactoring-inducing PRs? We firstly explored the presence of PRs that met our refactoring-inducing PR definition (Def. \ref{ripr}).
    \item RQ\textsubscript{2}: How do refactoring-inducing PRs compare to non-refactoring-inducing ones? We quantitatively investigated code reviewing-related aspects aiming to find out similarities/differences in PRs based on the refactorings performed.
    \item RQ\textsubscript{3}: Is refactoring induced by code reviews? We qualitatively scrutinized a stratified sample of refactoring-inducing PRs to validate the occurrence of refactoring edits induced by code reviewing, by manually examining review comments and discussions. 
\end{itemize}

\begin{figure*}[h]
    \centering
    \includegraphics[width=\textwidth]{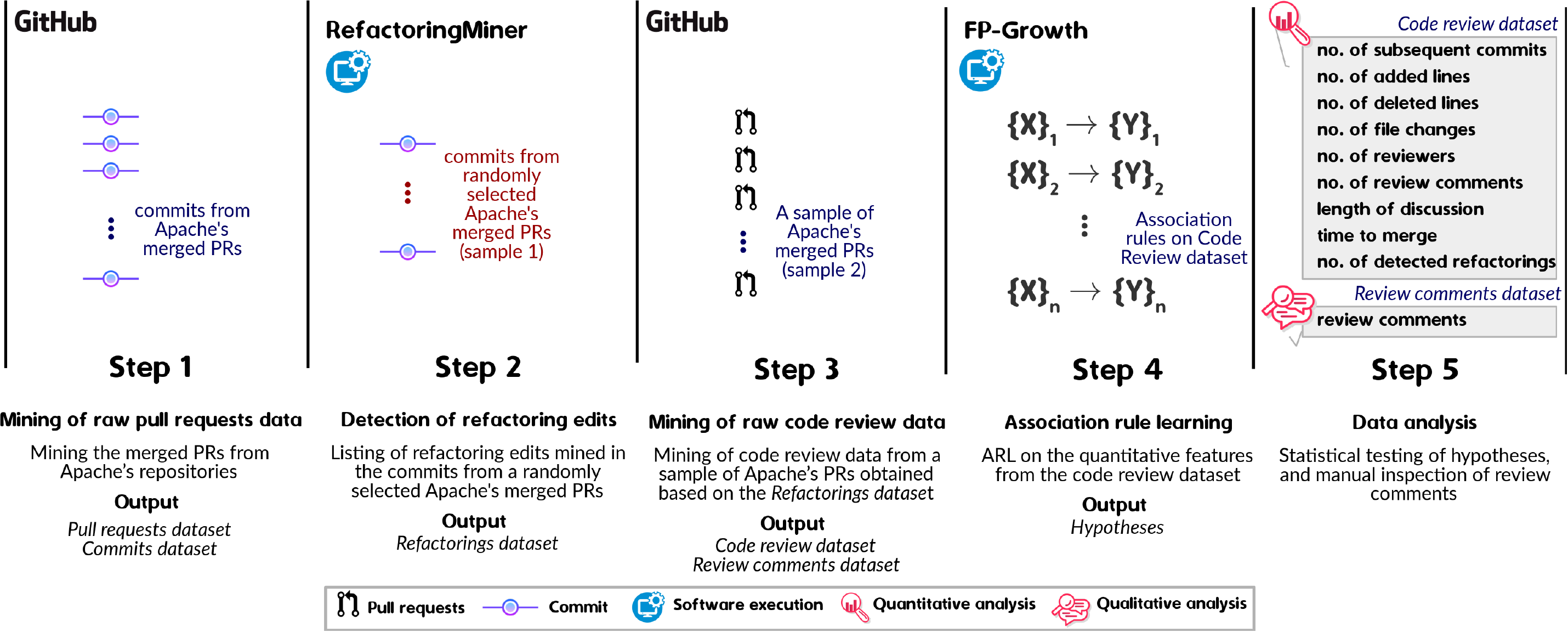}
	\vspace{-7mm}
	\caption{Overview of our Investigation.}
	\label{fig:CaseStudyDesignTwo}
\end{figure*}

Accordingly, supported by guidelines \cite{Runeson_ESE_2009}, we designed an empirical study that comprises five steps, as shown in Figure \ref{fig:CaseStudyDesignTwo} and described in the next subsections. Also, we made publicly available a reproduction kit containing the mined datasets and developed scripts for replicating the results for our research questions \cite{Reproduction_kit}.

\vspace{15mm}
\subsection{Mining Merged Pull Requests} 
\label{mining_merged_pull_requests}
We mined merged PRs from Apache's repositories at GitHub. We focused on merged PRs because they reveal actions that were in fact finalized, therefore, we can get a more in-depth understanding of refactoring-inducement.
We chose GitHub due to its popularity \cite{Octoverse} and to the mining resources available through extensive APIs -- GitHub REST API v3 \cite{GitHub_REST} and GitHub GraphQL API v4 \cite{GitHub_GRAPHQL}.

The \textit{Apache Software Foundation} (ASF) manages more than 350 open-source projects, with more than 8,000 contributors from all over the world; all of its projects migrated to GitHub in February 2019 \cite{Apache_Blog}.
Given Apache's popularity and relevance of contributions in the open-source software development context, we selected it for mining PRs \cite{Apache_Stats}. 
The refactoring mining tool we selected (Section \ref{refactoring-detection}) only supports projects developed in the Java, so we considered Java projects (almost 57\% of Apache's code is developed in Java).

In August 2019, we searched on Apache's non-archived Java repositories in GitHub (to take into account only actively maintained repositories), resulting in 65,006 merged PRs, detected in 467 out of 956 repositories; we then implemented a script to mine their merged PRs. 
We obtained two datasets: \textit{pull requests dataset} consists of 48,338 merged PRs (\textit{merge PR option}) from 453 distinct repositories while \textit{commits dataset} contains 53,915 recovered commits from 16,668 merged PRs (\textit{squash and merge} or \textit{rebase and merge} options) from 255 repositories.

\begin{figure}[h]
    \begin{center}
    \vspace{-1mm}
		{\includegraphics[scale = 0.4]{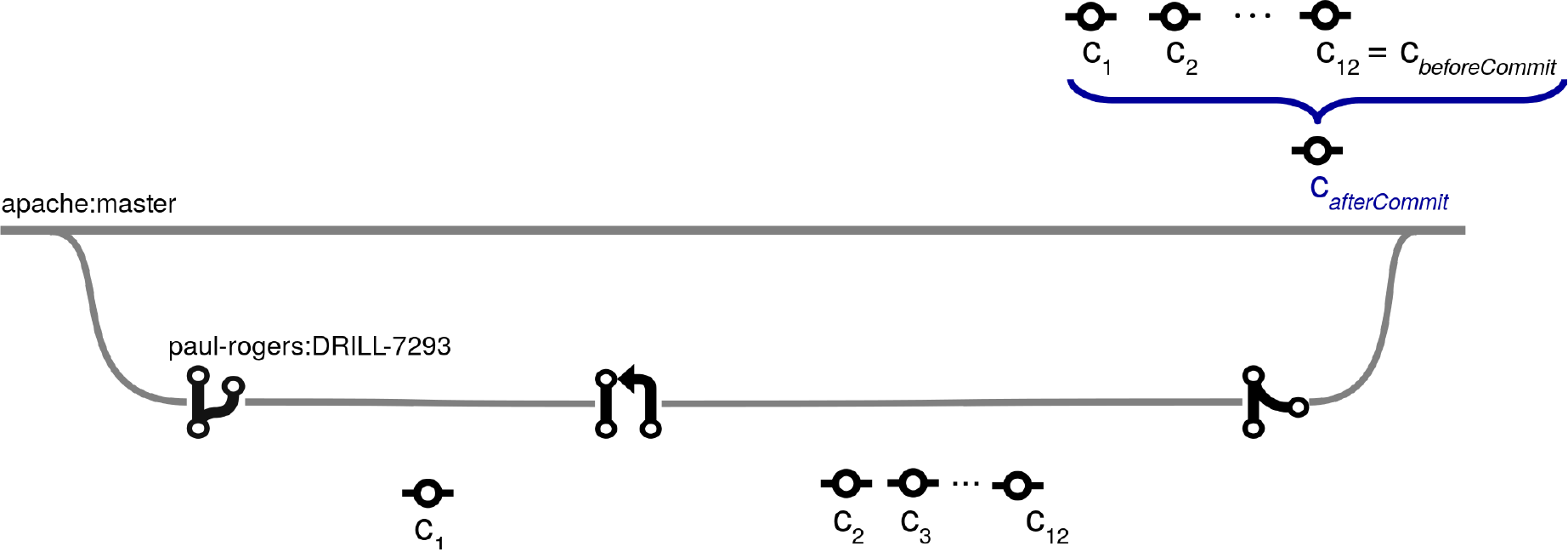}}
	\end{center}
	\vspace{-5mm}
	\caption{An Overview of Apache Drill PR \#1807, Illustrating Squashed Commits ($c_1-c_{12}$).}
	\vspace{-3mm}
	\label{fig:squashed-commits}
\end{figure}

Then, we recovered the commit history of squashed and merged PRs before any exploration of its original commits, assisted by the \textit{HeadRefForcePushedEvent} object accessible via GitHub GraphQL API  \cite{GitHub_GRAPHQL}. To clarify, consider the Apache's PR 1807 (Figure \ref{fig:squashed-commits}) that, originally, had 12 commits ($c_1-c_{12}$) that were squashed into single commit ($c_{afterCommit}$) after a \textit{force--pushed} event. Consequently, only one commit may be gathered from the PR ($c_{afterCommit}$).

Our recovery strategy follows two steps: (1) we recover the commits $c_{afterCommit}$ and $c_{beforeCommit}$ through \textit{HeadRefForcePushedEvent} object; and (2) we rebuild the original commits' history through tracking the commits from $c_{beforeCommit}$, which has the same value of $c_{12}$, until reaching the same SHA of the $c_{afterCommit}$'s parent, by using the \textit{compare} operation, as available in GitHub REST API v3 \cite{GitHub_REST}.
We executed the strategy's Step 1 for gathering the after and before commits from 65,006 pull requests, obtaining 53,915 commits after running the strategy's Step 2.

\begin{figure}[h]
      \centering
      \vspace{-2mm}
      \includegraphics[width=\linewidth]{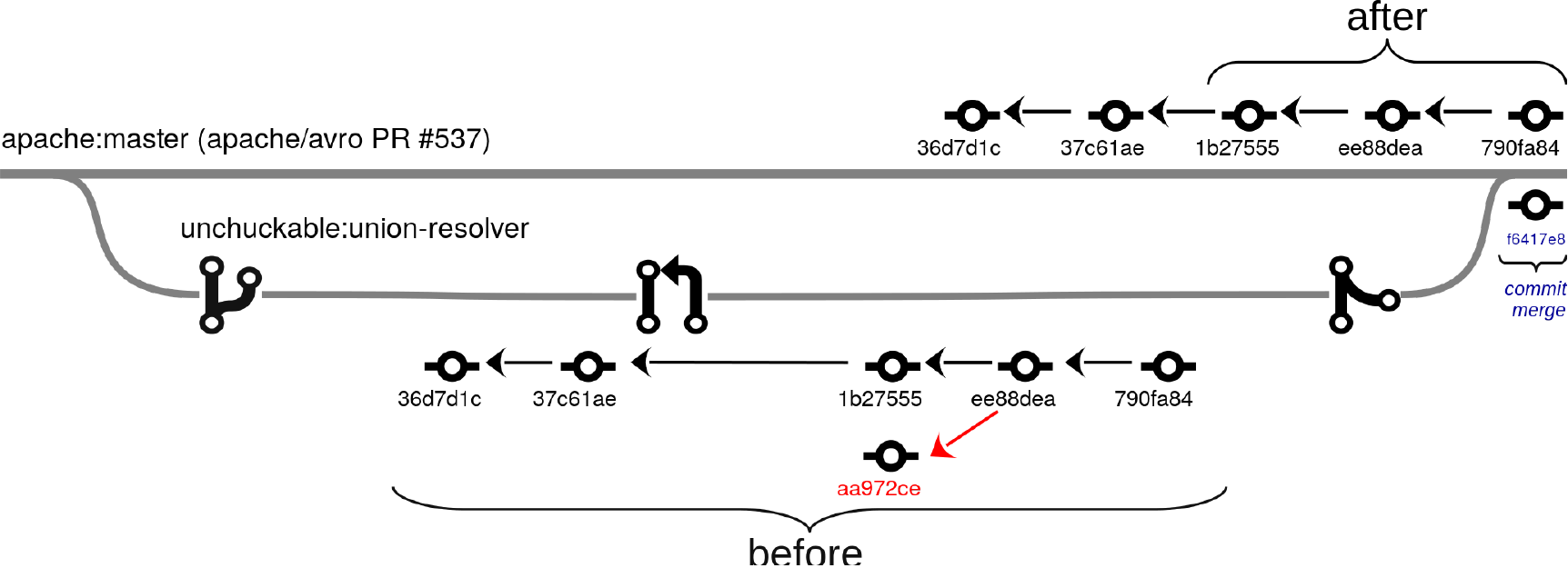}
      \vspace{-7mm}
      \caption{Illustrating a Pull Request's Commit Presenting Two Parents (Apache avro PR \#537)}
      \vspace{-3mm}
      \label{fig:commit-merge-problem}
    \end{figure}

We discarded PRs merged by \textit{rebase and merge} option since, in rebasing, some commits within the PR may be due to external changes (outside the scope of the code review sequence), conveying a threat to the validity, as argued in \cite{Paixao_SCAM_2019}. 
Accordingly, we considered the number of \textit{HeadRefForcePushedEvent} events and PR commits to identify PRs merged by \textit{squash and merge}. In specific, PRs merged by \textit{merge pull request} and \textit{squash and merge} present zero and one \textit{HeadRefForcePushedEvent} event, respectively (squashed and merged PRs keep one PR commit). Moreover, we dropped all PRs containing at least one subsequent commit with two parents, because such commits may represent external changes rebased onto a branch, as depicted in Figure \ref{fig:commit-merge-problem}. 
Note that, once commit \textit{ee88dea} has two parents, it integrates external changes, which were not reviewed in PR reviewing time.

\subsection{Refactoring Detection} 
\label{refactoring-detection}
RefactoringMiner detects refactorings in Java projects, presenting better results when compared to its competitors (precision of 99.6\% and recall of 94\%) \cite{Tsantalis_ICSE_2018, Tsantalis_TSE_2020}.
We considered version 2.0, which supports over 40 different refactoring types, including low-level refactorings, such as variable renames and extractions, allowing us to work with a more comprehensive list of refactoring edits.
For these reasons, we selected it for refactoring detection (Step 2).
In essence, it identifies the refactorings performed in a commit in relation to its parent commit, displaying a description of the applied refactorings (type and associated targets, e.g., the methods and classes involved in an \textit{Extract and Move Method} refactoring).
In this step, we considered only merged PRs containing two or more commits (\textit{sample 1}, Figure \ref{fig:CaseStudyDesignTwo}) intending to conform with our refactoring-inducing PR definition. 
After three weeks of RefactoringMiner running, we obtained a random sample of 225,127 detected refactorings in 8,761 merged PRs (13.5\% of the total number of Apache's merged PRs) from 209 distinct repositories, embracing 68,209 commits.
The source of randomness lies in the order in which the repositories were processed.

At that point, we checked the \textit{commits' authored date} against the \textit{PRs' opening date} in order to identify initial and subsequent commits for the sample's PRs. Therefore, the number of refactorings of a PR takes into account only subsequent commits.

\subsection{Mining Code Review Data} 
Empirical studies have investigated code review efficiency and effectiveness to understand the practice, elaborate recommendations, and develop improvements. 
Together, these studies share a set of useful code review aspects for further investigation, such as change description \cite{Bosu_MSR_2015,Tao_FSE_2012}, code churn \cite{Thongtanunam_ESE_2017}, length of discussion \cite{Kononenko_ICSE_2016,McIntosh_MSR_2014,Rigby_FSE_2013,Thongtanunam_ESE_2017}, number of changed files \cite{Bosu_MSR_2015,Kononenko_ICSE_2016}, number of commits \cite{McIntosh_MSR_2014,Rigby_SEM_2014}, number of people in the discussion \cite{Kononenko_ICSE_2016}, number of resubmissions \cite{Kononenko_ICSE_2016,Rigby_FSE_2013}, number of review comments, \cite{Beller_MSR_2014,McIntosh_MSR_2014,Rigby_FSE_2013}, number of reviewers \cite{Rigby_FSE_2013,Sauer_TSE_2000}, size of change \cite{Baysal_ESE_2016,Kononenko_ICSE_2016,Rigby_FSE_2013}, and time to merge \cite{Gousios_ICSE_2014,Izquierdo_MSR_2016}. Therefore, the mining of raw code review data (Step 3) consisted of collecting the code reviewing-related attributes listed in Table \ref{tab:variables}, considering 8,761 PRs from Step 2 (\textit{sample 2}, Figure \ref{fig:CaseStudyDesignTwo}).
Attributes number, title, labels, and repository's name are useful to uniquely identify a PR.
We clarify that we do not count the distinct files changed (i.e., the set of the changed files), but the number of times the files changed (i.e., the list of file changes) over subsequent PR commits. Hence, the number of added lines and deleted lines denote the number of lines modified across file changes.    

\begin{table}
    \caption{Selected Pull Request Attributes for Mining}
    \vspace{-5mm}
    \begin{center}
    \begin{tabular}{cl}
        \toprule
        \textit{\textbf{\small{Attribute}}} &  \textit{\textbf{\small{Description}}}\\
        \midrule
        \small{number} & 
            \small{Numerical identifier of a PR} \\ 
        \hline
        \small{title} & 
            \small{Title of a PR} \\ 
        \hline
        \small{repository} &
            \small{Repository's name of a PR} \\
        \hline
        \small{labels} & 
            \small{Labels associated with a PR} \\ 
        \hline
        \small{commits} &
            \small{No. of subsequent commits in a PR} 
        \\ \hline
        \small{additions} &
            \small{No. of added lines in a PR} 
        \\ \hline
        \small{deletions} & 
            \small{No. of deleted lines in a PR} \\
        \hline
        \small{file changes} &
            \small{No. of file changes in a PR}
        \\ \hline
        \small{creation date} & 
            \small{Date and time of a PR creation} \\
        \hline
        \small{merge date} &  
            \small{Date and time of a PR merge} \\ 
        \hline
        \small{review comments} & 
            \small{No. of review comments in a PR} 
        \\ \hline
        \small{non-review comments} & 
            \small{No. of non-review comments in a PR} \\
        \bottomrule
    \end{tabular}
    \end{center}
    \label{tab:variables}
    \vspace{-3mm}
\end{table}

For mining, we imposed one \textit{precondition}: only merged PRs, comprising at least one review comment, should be mined aiming to explore refactoring-inducement and to collect review comments for further investigation.
Thus, the mining generated two datasets, \textit{code review dataset} and \textit{review comments dataset}, refined according to the following procedures: dropping merged PRs with inconsistencies, such as zero file changes and zero reviewers; checking for duplicates; and mining from non-mirrored repositories.
As a result, our final sample consists of code review data from 1,845 merged PRs (2.8\% of the total number of Apache's merged PRs from Step 1 and 21.1\% of the number of sample's PRs obtained from Step 2), encompassing 4,702 subsequent commits, 6,556 detected refactorings, and 12,547 review comments, mined from 84 distinct Apache's repositories. 

\subsection{Association Rule Learning}\label{arl-design}
Aiming to explore what differentiates refactoring-inducing PRs from non-refactoring-inducing ones, we executed ARL (Step 4). 
Such strategy assists exploratory analysis by identifying natural structures derived from the relationships between the characteristics of data \cite{Celebi_Springer_2016}.
Accordingly, by considering ARL on refactoring-inducing and non-refactoring-inducing PRs, we can identify ARs that likely support us in the formulation of more accurate hypotheses concerning differences/similarities between those two groups.
One may argue that clustering is a better alternative than ARL to find groups of PRs with distinct characteristics. Nonetheless, we experimentally performed clustering in our sample of PRs, after conducting a rigorous selection of clustering algorithm and input parameters\footnote{\textit{We used the Ordering Points To Identify the Clustering Structure} (OPTICS) algorithm \cite{Ankerst_SIGMOD_1999} and \textit{Euclidean distance} \cite{LinearAlgebra_Book_2013} as similarity metric.}, but we found a great noise ratio (76.3\%).

\subsubsection{Selection of features}
We selected all features that can be represented as a number regarding changes, code review, and refactorings, from the code review dataset (Step 3). 
We considered a three-context perspective (changes, code review, and refactorings) because they together might potentially support the identification of differences between refactoring-inducing and non-refactoring-inducing PRs.
These are the selected features: number of subsequent commits, number of file changes, number of added lines, number of deleted lines, number of reviewers, number of review comments, length of discussion, time to merge, and number of detected refactorings.
Note that the length of discussion and time to merge are derived from \textit{review comments} $+$ \textit{non-review comments}, and \textit{merge date} $-$ \textit{creation date} (in number of days), respectively.

One may argue that other features could also be considered; however, (i) the PR title is written using natural language, so it is subject to ambiguities; (ii) PR labels are not mandatory, only 349 PRs from our sample have labels; (iii) date and time of creation/merge are specific values, so we used the difference between them (time to merge) for exploration; and (iv) the number of non-review comments of a PR is part of its length of discussion.

\subsubsection{Feature engineering}
We applied one-hot encoding based on the quartiles of the features, resulting in the binning presented in Table \ref{tab:onehot}. We chose such technique due to its simplicity and linear time and space complexities \cite{MachineLearning_Book_2018}.
We did not discard the outliers because, in the context of this study, they do not represent experimental errors; thus, they can potentially indicate circumstances for further examination. Consequently, the \textit{very high} category (fourth quartile) includes the outliers.

\begin{table}
    \caption{One-Hot Encoding for Binning of Features}
    \label{tab:onehot}
    \vspace{-5mm}
    \begin{center}
    \begin{tabular}{cc}
        \toprule
        \textit{\textbf{\small{Category}}} &  \textit{\textbf{\small{Range}}}\\
        \midrule
        None & 0 \\\hline
        Low & $0 < quantile \leq 0.25$ \\\hline
        Medium & $ 0.25 < quantile \leq 0.50$ \\\hline
        High & $0.50 < quantile \leq 0.75$ \\\hline
        Very high & $0.75 < quantile \leq 1.0$\\
        \bottomrule
    \end{tabular}
    \end{center}
    \vspace{-3mm}
\end{table}

\subsubsection{Selection and execution of an algorithm}
We selected the FP-Growth algorithm due to its performance \cite{Han_SIGMOD_2000}.
Then, we developed a script for the ARL by using the FP-growth implementation available in the \textit{mlxtend.frequent\_patterns} module \cite{Raschka_OSS_2018}.
We set the minimum support threshold to 0.1 to avoid discarding likely ARs for further analysis \cite{Coenen_DMKD_2004}.
Aiming to get meaningful ARs, we considered minimum thresholds for confidence $\geq$ 0.5, lift $>$ 1, and conviction $>$ 1. 
We performed a prior experiment concerning values of minimum support and minimum confidence by taking the thresholds considered in \cite{Agrawal_SR_1993} as a reference (support of 0.01, confidence of 0.5). 
We ran FP-growth considering support values ranging from 0.01 to 0.1 by steps of 0.01, and confidence 0.5 (Table \ref{tab:supp-conf}). 
In all these settings, we found ARs that cover all input features. Since support is a statistical significance measure, we consider the last setting (minimum support of 0.1, confidence of 0.5) for purposes of FP-growth execution.
A lift threshold $>$ 1 reveals useful ARs \cite{Berzal_IDA_2002}, while a conviction threshold $>$ 1 denotes ARs with logical implications \cite{Brin_CMD_1997}.

\begin{table}
    \caption{ARL Output by Experimenting Minimum Support from 0.01 to 0.1 by Steps of 0.01, and Confidence of 0.5}
    \vspace{-5mm}
    \begin{center}
    \begin{tabular}{cc}
        \toprule
        \textit{\textbf{\small{Support $\geq$}}} &  \textit{\textbf{\small{Number of association rules}}}\\
        \midrule
        0.01 & 52,944\\\hline
        0.02 & 19,239\\\hline
        0.03 & 10,354\\\hline
        0.04 & 5,567\\\hline
        0.05 & 3,572\\\hline
        0.06 & 2,264\\\hline
        0.07 & 1,640\\\hline
        0.08 & 1,004\\\hline
        0.09 & 712\\\hline
        0.10 & 562\\
        \bottomrule
    \end{tabular}
    \end{center}
    \label{tab:supp-conf}
    \vspace{-3mm}
\end{table}

\subsubsection{Interpretation of results}
We considered the feature levels (\textit{none}, \textit{low}, \textit{medium}, \textit{high}, and \textit{very high}), instead of absolute values, as items for composing ARs aiming to identify relative associations among two groups for investigation, e.g., \{\textit{high number of added lines}\} $\rightarrow$ \{\textit{high number of reviewers}\}.
The ARs work as basis for the formulation of hypotheses regarding the characterization of our sample's PRs. 
In this sense, we carried out the following procedure:

\begin{enumerate}[leftmargin=*]
    \item manual examination of the ARs to recognize potential differences/similarities that support the formulation of hypotheses;
    \item analysis of the pairwise ARs, ARs containing the \textit{number of refactorings} as an item, and ARs whose conviction is infinite to assist the rationale for the formulation of hypotheses; and
    \item formulation of hypotheses to quantitatively investigate the differences between refactoring-inducing and non-refactoring-inducing PRs.
\end{enumerate}

\subsection{Data Analysis}
\subsubsection{Quantitative data analysis}
We analyzed the output of Step 3 by exploring the detected refactorings by PR to answer RQ${_1}$. The number of refactorings was computed by considering the edits detected as in the PR \textit{subsequent} commit(s).
As a complement, we computed a 95\% confidence interval for the percentual (proportion) of refactoring-inducing PRs in Apache's merged PRs, by performing \textit{bootstrap resampling} \cite{Book_Bootstrap_1993}. 
We applied statistical testing of hypotheses intending to answer RQ${_2}$. 
That analysis encompassed the testing of eight hypotheses formulated from the analysis of the ARL output (Step 4), driven by a comparison between refactoring-inducing and non-refactoring-inducing PRs.
We executed each hypothesis testing in line with this workflow, guided by \cite{Stats_Book_2010}:

\begin{enumerate}[leftmargin=*]
    \item Definition of null and alternative hypotheses.
    \item Performing of statistical test. We considered a significance level of 5\%, and a substantive significance (effect size) for denoting the magnitude of the differences between refactoring-inducing and non-refactoring-inducing PRs at the population level. First, we checked the assumptions for parametric statistical tests (steps \textit{a} and \textit{b}), since the independence assumption is already met (i.e., a PR is either a refactoring-inducing or not). For exploring the difference between refactoring-inducing and non-refactoring-inducing PRs, we computed a 95\% confidence interval by bootstrapping resample according to the output from steps \textit{a} and \textit{b}, in mean or median (step \textit{c}). Then, we conducted a proper statistical test and calculated the effect size (step \textit{d}). 
    \begin{enumerate}[leftmargin=*]
        \item checking for data normality by using the \textit{Shapiro-Wilk} test;
        \item checking for homogeneity of variances via \textit{Levene's} test;
        \item computation of confidence interval for the difference in mean or median aligned to output from steps \textit{a} and \textit{b};
        \item performing of either parametric independent \textit{t-test} and \textit{Cohen's d}, or non-parametric \textit{Mann Whitney U} test and \textit{Common-Language Effect Size} (CLES) in line with the output from steps \textit{a} and \textit{b}. CLES is the probability, at the population level, that a randomly selected observation from a sample will be higher/greater than a randomly selected observation from another sample \cite{McGraw_CLES_1992}.
    \end{enumerate} 
    \item Deciding if the null hypothesis is supported or refused.
    \newline
\end{enumerate}

\vspace{-5mm}
\subsubsection{Qualitative data analysis} \label{qualitative-data-analysis-design} 
In order to answer RQ$_3$, three developers (intending to mitigate researcher bias) manually examined review discussions and validated the detected refactorings from a subset of refactoring-inducing PRs of our sample. 
We adopted a stratified random sampling to select refactoring-inducing PRs for an in-depth investigation of their review comments and discussion while cross-referencing their detected refactoring edits. Moreover, we validated these refactorings by checking for false positives. As a whole, the qualitative analysis lasted 30 days.  
We chose that sampling strategy because it provides a means to sample non-overlapping subgroups based on specific characteristics \cite{Marshall_FP_1996}, (e.g. number of refactorings), where each subgroup (\textit{stratum}) can be sampled using another sampling method -- a setting that quite fits to further investigation of categories of refactoring-inducing PRs containing a \textit{low}, \textit{medium}, \textit{high}, and \textit{very high} number of refactorings (Table \ref{tab:onehot}).
To define the sample size, we considered a confidence level of 95\% and a margin of error of 5\%, so obtaining 228, thus considering 57 refactoring-inducing PRs randomly selected from each category.
We split the samples into four categories based on the numbers of refactorings in order to check if there is a difference in the effect of code review refactoring requests/inducement between PRs with massive refactoring efforts versus PRs with small/focused refactoring efforts.

In the analysis, firstly, we conducted a calibration in which one of the analysts followed up ten analyses performed by the others. Next, each analyst apart examined 40.3\%, 38.2\%, and 21.5\% of the data.
In such subjective decision-making, we considered the refactoring-inducement in settings where review comments either explicitly suggested refactoring edits (e.g., ``How about renaming to ...?''\footnote{Apache samza PR \#1051, available in \url{https://git.io/J3z9H}.}) or left any actionable recommendation that induced refactoring (e.g., ``avoid multiple booleans'' induced a \textit{Merge Parameter} instance\footnote{Apache fluo PR \#1032, available in \url{https://git.io/J3mxZ}.}).

\section{Results and Discussion} \label{results-and-discussion}
\subsection{How Common are Refactoring-Inducing Pull Requests?}
\begin{figure}[h]
    \vspace{-2mm}
    \begin{subfigure}{.235\textwidth}
      \centering
      \includegraphics[width=\linewidth]{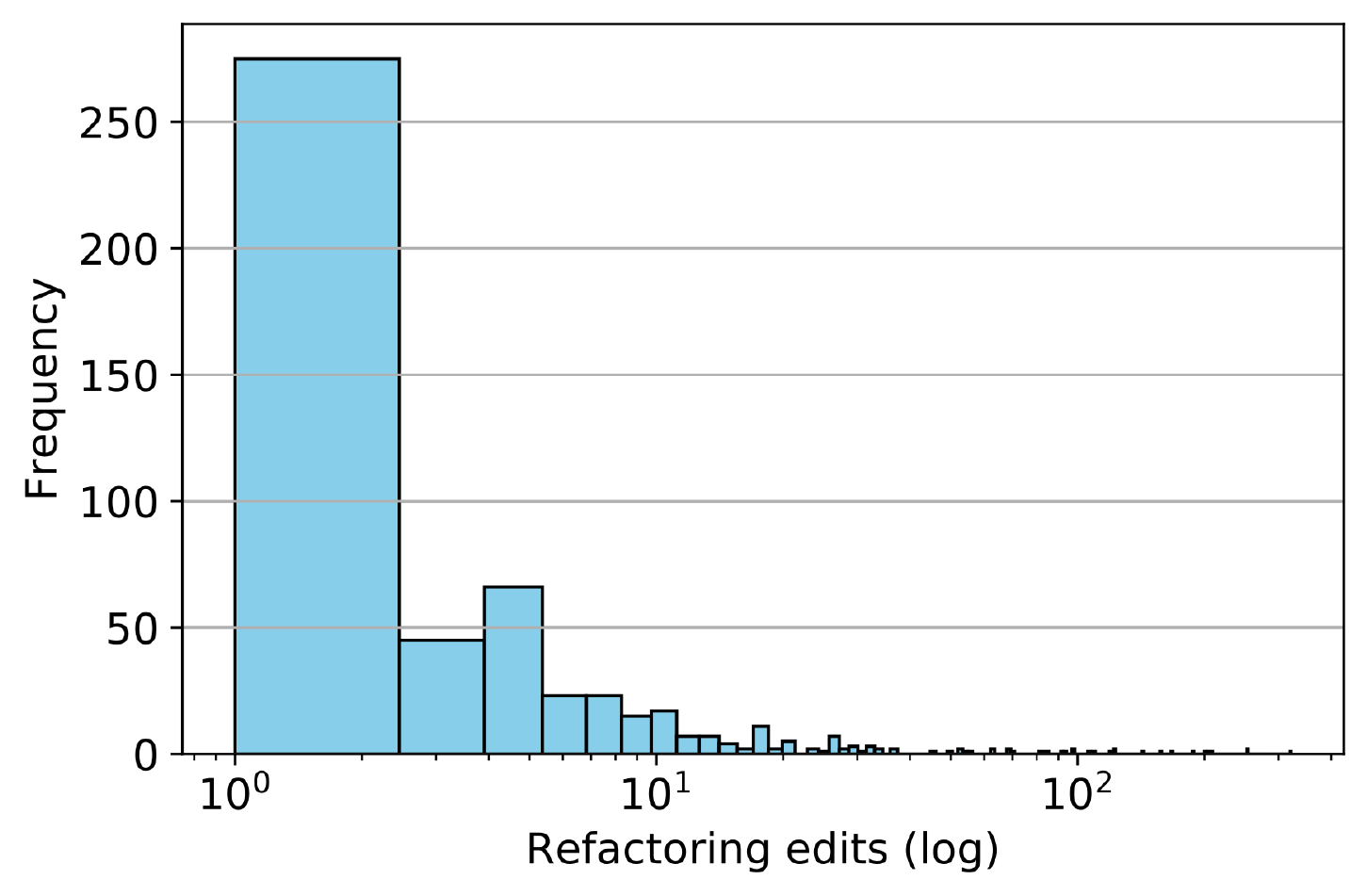}
      \vspace{-5mm}
      \caption{Histogram}
	  \label{fig:n_refactorings_histogram}
    \end{subfigure}
    \begin{subfigure}{.235\textwidth}
      \centering
      \includegraphics[width=\linewidth]{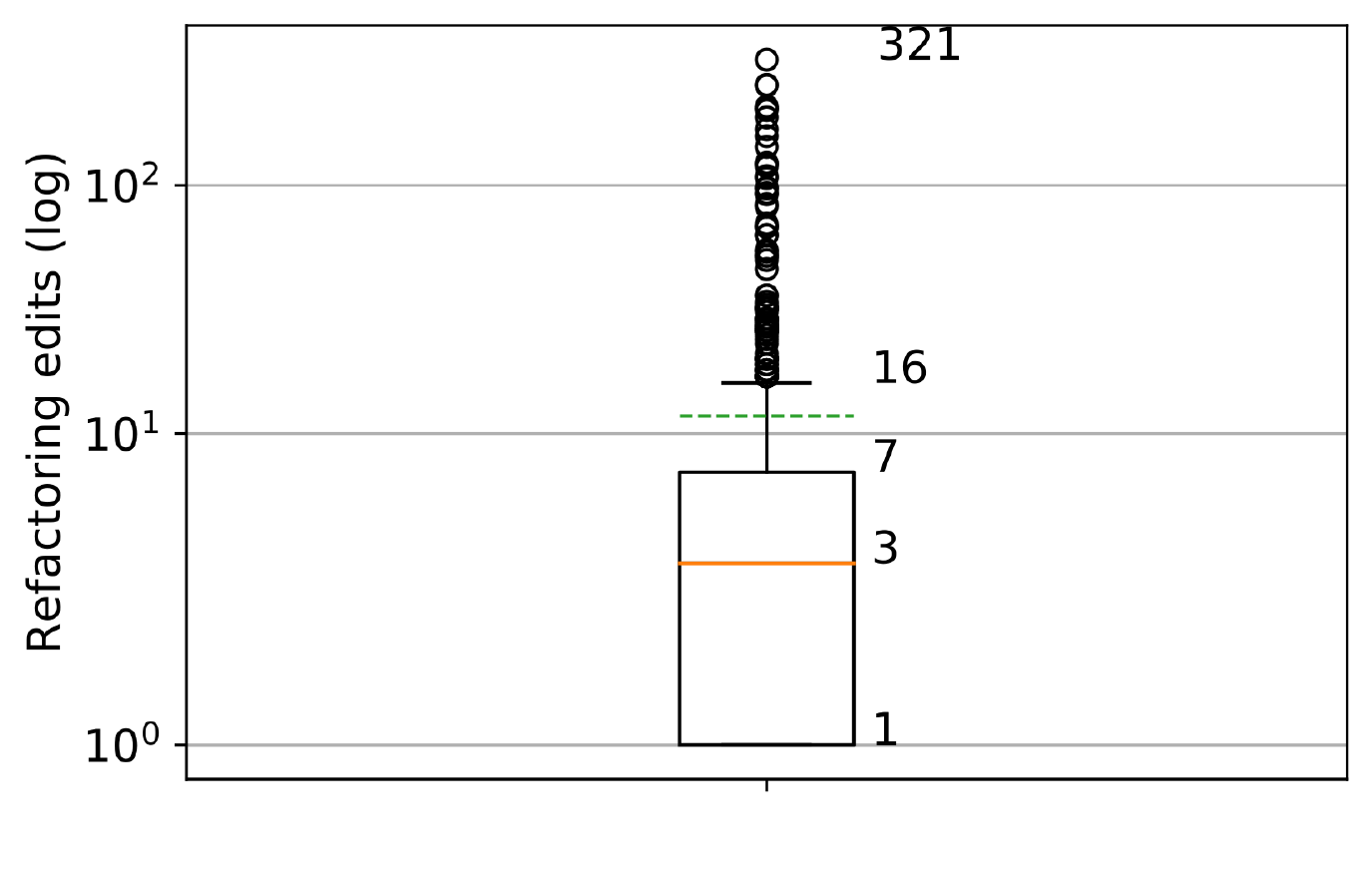}
      \vspace{-5mm}
      \caption{Boxplot}
	  \label{fig:n_refactorings_boxplot}
    \end{subfigure}
    \vspace{-4mm}
    \caption{Refactorings in the Refactoring-Inducing PRs}
	\label{fig:n_refactorings}
\end{figure}

We found 557 refactoring-inducing PRs (30.2\% of our sample's PRs), equaling 12,547 detected refactoring edits.
As shown in Figure \ref{fig:n_refactorings_histogram}, the histogram of refactoring edits is positively skewed, presenting outliers. Thus, a low number of refactoring edits is quite frequent.
The number of refactorings by PR is 11.8 on average (SD = 32.3) and 3 as median (IQR = 6), according to Figure \ref{fig:n_refactorings_boxplot}. 

Using bootstrapping resampling and a 95\% confidence level, we obtained a confidence interval ranging from 28.1\% to 32.3\% for the proportion of refactoring-inducing PRs in Apache's merged PRs.
These results reveal significant refactoring activity induced in PRs. 
This is a motivating result, while outliers' presence can indicate scenarios scientifically relevant for further exploration.

\begin{tcolorbox}[colback = white, left=0pt, right=0pt, top=0pt, bottom=0pt]
\textbf{Finding 1}: We found 30.2\% of refactoring-inducing PRs, which percentage (proportion) in Apache's merged PRs is in [28.1\%, 32.3\%], for a 95\% confidence level.
\end{tcolorbox}
\vspace{-3mm}

\subsection{How Do Refactoring-Inducing Pull Requests Compare to non-Refactoring-Inducing Ones?}
From ARL, we obtained 562 ARs (146 from refactoring-inducing PRs and 416 from non-refactoring-inducing PRs).
Then, we manually inspected them, by searching for pairwise ARs (AR$_1$--AR$_{7}$), ARs whose conviction is infinite (AR$_5$, AR$_6$), and the remaining ARs (AR$_{2}$, AR$_{3}$, AR$_{4}$). Accordingly, we selected four ARs (AR$_1$--AR$_{4}$) obtained from refactoring-inducing PRs and three ARs (AR$_5$--AR$_{7}$) from non-refactoring-inducing PRs, all catalogued in Table \ref{tab:associations_from_inspection}, in decreasing order of conviction.
Since we did not identify the same pairs of ARs in both groups, we needed to consider a distinct number of ARs (hence, itemsets) for the comparison purpose when addressing all features.
Afterwards, we carried out an analysis of those ARs. We formulated eight hypotheses on the differences/similarities between refactoring-inducing and non-refactoring-inducing PRs, discussed as follows. Table \ref{tab:statistics_prs} shows the \textit{average}, \textit{Standard Deviation} (SD), \textit{median}, and \textit{Interquartile Range} (IQR) of the examined features from refactoring-inducing and non-refactoring-inducing PRs.

\begin{table*}
    \caption{Association Rules Selected by Manual Inspection (AR$_1$--AR$_{4}$ for Refactoring-Inducing PRs, AR$_5$--AR$_{7}$ for non-Refactoring-Inducing PRs)}
    \vspace{-5mm}
    \begin{center}
    \begin{tabular}{clcccc}
        \toprule
        \multicolumn{1}{c}{\textbf{\small \textit{Id}}} & \multicolumn{1}{c}{\textbf{\small \textit{Association rule}}} &
        \multicolumn{1}{c}{\textbf{\small \textit{Supp}}} & \multicolumn{1}{c}{\textbf{\small \textit{Conf}}} &
        \multicolumn{1}{c}{\textbf{\small \textit{Lift}}} &
        \multicolumn{1}{c}{\textbf{\small \textit{Conv}}} \\ \midrule
         AR$_1$ &\small\begin{tabular}[l]{@{}l@{}} 
            \{very high length of discussion, very no. of reviewers\} $\rightarrow$ \{very high no. of review comments\}\\
            \end{tabular}
            & 0.13 & 0.85 & 3.08 & 4.89\\\hline
            
        AR$_2$ &\small\begin{tabular}[l]{@{}l@{}} 
            \{very high no. of added lines, very high no. of subsequent commits\} $\rightarrow$ \{very high no. of file changes\}\\
            \end{tabular}
            & 0.11 & 0.83 & 3.23 & 4.51\\\hline 
            
        AR$_3$ &\small\begin{tabular}[l]{@{}l@{}} 
            \{very high no. of deleted lines, very high no. of subsequent commits\} $\rightarrow$ \{very high no. of file changes\}\\
            \end{tabular}
            & 0.10 & 0.81 & 3.12 & 3.81\\\hline  
            
        AR$_4$ &\small\begin{tabular}[l]{@{}l@{}} 
            \{medium time to merge\} $\rightarrow$ \{very high no. of reviewers\}\\
            \end{tabular}
            & 0.16 & 0.51 & 1.06 & 1.06\\
            
        \midrule[1pt]
             
        AR$_5$ &\small\begin{tabular}[l]{@{}l@{}} 
            \{high no. of subsequent commits, low no. of added lines, low no. of deleted lines\} $\rightarrow$ \{medium no. of file\\ changes\}\\
            \end{tabular}
            & 0.12 & 1.00 & 2.63 & $\infty$\\\hline
            
        AR$_{6}$ &\small\begin{tabular}[l]{@{}l@{}} 
            \{medium no. of file changes, very high no. of reviewers, medium time to merge\} $\rightarrow$ \{high no. of \\subsequent commits\}\\
            \end{tabular}
            & 0.13 & 1.00 & 1.83 & $\infty$\\\hline
            
        AR$_{7}$ &\small\begin{tabular}[l]{@{}l@{}} 
            \{very high no. of reviewers, medium length of discussion\} $\rightarrow$ \{medium no. of review comments\}\\
            \end{tabular}
            & 0.13 & 0.61 & 1.71 & 1.63 \\   

        \bottomrule
    \end{tabular}
    \end{center}
    \label{tab:associations_from_inspection}
\end{table*}

\begin{table*}
    \vspace{-3mm}
    \caption{Statistics of the Pull Requests Attributes}
    \vspace{-5mm}
    \begin{center}
    \begin{tabular}{l c c c c c c c c}
    \toprule
    \multirow{2}{*}{
        \begin{tabular}[c]{@{}c@{}}\textbf{\textit{\small Pull Request Attribute}}\end{tabular}} &
    \multicolumn{4}{c}{\textbf{\textit{\small Refactoring-Inducing Pull Requests}}} & \multicolumn{4}{c}{\textbf{\textit{\small non-Refactoring-Inducing Pull Requests}}}\\\cline{2-9} &
        \multicolumn{1}{c}{\small Average} & 
        \multicolumn{1}{c}{\small SD} & 
        \multicolumn{1}{c}{\small Median} & 
        \multicolumn{1}{c}{\small IQR} & 
        \multicolumn{1}{c}{\small Average} & 
        \multicolumn{1}{c}{\small SD} & 
        \multicolumn{1}{c}{\small Median} & 
        \multicolumn{1}{c}{\small IQR} \\
        \midrule
        
        \multicolumn{1}{l}{\textit{\small Number of added lines}} & 
            \multicolumn{1}{c}{945.9} & 4,744.3 & 72 & 250 & \multicolumn{1}{c}{57.5} & 517.8 & 8 & 28 \\\hline
        
        \multicolumn{1}{l}{\textit{\small Number of deleted lines}} &
            \multicolumn{1}{c}{377.4} & 1,859.7 & 41 & 139 &
            \multicolumn{1}{c}{41.2} & 303.8 & 6 & 16.2\\\hline
        
        \multicolumn{1}{l}{\textit{\small Number of file changes}} & 
            \multicolumn{1}{c}{ 32.1} & 119.7 & 7 & 15 &
            \multicolumn{1}{c}{6.1} & 60.2 & 2 & 3\\\hline
    
        \multicolumn{1}{l}{\textit{\small Number of subsequent commits}} & 
            \multicolumn{1}{c}{3.7} & 3.4 & 3 & 2 &
            \multicolumn{1}{c}{2.1} & 1.9 & 1 & 1\\\hline
   
        \multicolumn{1}{l}{\textit{\small Number of review comments}} & 
            \multicolumn{1}{c}{9.8} & 11.1 & 6 & 9 &
            \multicolumn{1}{c}{5.5} & 8.2 & 3.5 & 4\\\hline
          
        \multicolumn{1}{l}{\textit{\small Length of discussion}} & 
            \multicolumn{1}{c}{15.2} & 13.8 & 11 & 14 &
            \multicolumn{1}{c}{10.1} & 12.1 & 7 & 8\\\hline
            
        \multicolumn{1}{l}{\textit{\small Number of reviewers}} & 
            \multicolumn{1}{c}{2.3} & 0.9 & 2 & 1 &
            \multicolumn{1}{c}{2.1} & 0.9 & 2 & 0\\\hline
       
        \multicolumn{1}{l}{\textit{\small Time to merge (days)}} & 
            \multicolumn{1}{c}{14.3} & 45.6 & 5 & 11 &
            \multicolumn{1}{c}{9.3} & 33.1 & 2 & 7\\
    
        \bottomrule
    \end{tabular}
    \end{center}
    \label{tab:statistics_prs}
    \vspace{-3mm}
\end{table*}

\subsubsection*{\textbf{H\textsubscript{1}}} \textbf{Refactoring-inducing PRs are more likely to have more added lines than non-refactoring-inducing PRs} {\small{(AR$_2$/AR$_{3}$,AR$_{5}$)}}. 

\subsubsection*{\textbf{H\textsubscript{2}}} \textbf{Refactoring-inducing PRs are more likely to have more deleted lines than non-refactoring-inducing PRs} {\small{(AR$_2$/AR$_{3}$,AR$_{5}$)}}. 

\begin{tcolorbox}[colback = white, left=0pt, right=0pt, top=0pt, bottom=0pt]
\textbf{Finding 2}: Refactoring-inducing PRs comprise significantly more code churn than non-refactoring-inducing ones, since refactoring-inducing PRs are significantly more likely to have a higher number of added lines (U = $0.58 \times e^{+06}$, p $<$ .05), CLES = 81.2\% and deleted lines (U = $0.57 \times e^{+06}$, p $< $ .05), CLES = 80.5\% than non-refactoring-inducing PRs.
\end{tcolorbox}

This is an expected result in light of the findings from Heged\"us et al., since refactored code has significantly higher size-related metrics \cite{Hegedus_IST_2018}. 
We speculate that reviewing larger code churn may potentially promote refactorings. This understanding is supported by Rigby et al., who observed that the code churn's magnitude influences code reviewing \cite{Rigby_ICSE_2008,Rigby_SEM_2014}, and Beller et al. who discovered that the larger the churn, the more changes could follow \cite{Beller_MSR_2014}.

\subsubsection*{\textbf{H\textsubscript{3}}}
\textbf{Refactoring-inducing PRs are more likely to have more file changes than non-refactoring-inducing PRs} {\small{(AR$_2$/AR$_{3}$,AR$_{5}$)}}.

\begin{tcolorbox}[colback = white, left=0pt, right=0pt, top=0pt, bottom=0pt]
\textbf{Finding 3}: Refactoring-inducing PRs encompass significantly more file changes than non-refactoring-inducing ones (U = $0.56 \times e^{+06}$, p $< $ .05), CLES = 79.1\%.
\end{tcolorbox}

We conjecture that reviewing code arranged across files may motivate refactorings, an argument supported by Beller et al. regarding more file changes comprise more changes during code review \cite{Beller_MSR_2014}. 
By observing change-related aspects (churn and file changes), our findings confirm previous conclusions on the influence of the amount and magnitude of changes on code review \cite{Baysal_ESE_2016,Kononenko_ICSE_2016,Rigby_SEM_2014,Rigby_ICSE_2008}. 
When analyzing the changes and refactorings, our findings reinforce prior conclusions on refactored code significantly present higher size-related metrics (e.g., number of code lines and file changes) \cite{Hegedus_IST_2018}, and larger changes promote refactorings \cite{Paixao_TSE_2019}.

\subsubsection*{\textbf{H\textsubscript{4}}}
\textbf{Refactoring-inducing PRs are more likely to have more subsequent commits than non-refactoring-inducing PRs}\\ {\small{(AR$_{2}$/AR$_{3}$, AR$_{5}$)}}.  

\begin{tcolorbox}[colback = white, left=0pt, right=0pt, top=0pt, bottom=0pt]
\textbf{Finding 4}: Refactoring-inducing PRs comprise significantly more subsequent commits than non-refactoring-inducing PRs (U = $0.51 \times e^{+06}$, p $<$ .05), CLES = 70.6\%. 
\end{tcolorbox}

Based on our previous findings on the magnitude of code churn and file changes, that result is expected and aligned to Beller et al. concerning the impacts of larger code churn and wide-spread changes across files on consequent changes \cite{Beller_MSR_2014}. Accordingly, we speculate that reviewing refactoring-inducing PRs might require more subsequent changes, in turn, denoted by more subsequent commits in comparison with non-refactoring-inducing PRs. 

\subsubsection*{\textbf{H\textsubscript{5}}}
\textbf{Refactoring-inducing PRs are more likely to have more review comments than non-refactoring-inducing PRs} {\small{(AR$_{1}$, AR$_{7}$)}}.  

\begin{tcolorbox}[colback = white, left=0pt, right=0pt, top=0pt, bottom=0pt]
\textbf{Finding 5}: Refactoring-inducing PRs embrace significantly more review comments than non-refactoring-inducing PRs (U = $0.47 \times e^{+06}$, p $<$ .05), CLES = 65.1\%.
\end{tcolorbox}
 
Beller et al. found that the most changes during code review are driven by review comments \cite{Beller_MSR_2014}, and Pantiuchina et al. discovered that almost 35\% of refactoring edits are motivated by discussion among developers in OSS projects at GitHub \cite{Pantiuchina_SEM_2020}.
Thus, we conjecture that, besides change-related aspects, GitHub's PR model can constitute a peculiar structure for code review, in which review comments influence the occurrence of refactorings, therefore explaining our result. 
This argument originates from the fact that a pull-based collaboration workflow provides reviewing resources \cite{GitHub_PR} (e.g., a proper code reviewing UI) for developers to improve/fix the code while having access to the history of commits and discussion. 
Our finding also provides insight for examination of review comments to get an in-depth understanding of refactoring-inducement.

\subsubsection*{\textbf{H\textsubscript{6}}}
\textbf{Refactoring-inducing PRs are more likely to present a lengthier discussion than non-refactoring-inducing PRs} \\{\small{(AR$_{1}$, AR$_{7}$)}}.

\begin{tcolorbox}[colback = white, left=0pt, right=0pt, top=0pt, bottom=0pt]
\textbf{Finding 6}: Refactoring-inducing PRs enclose significantly more discussion than non-refactoring-inducing PRs (U = $0.46 \times e^{+06}$, p $< $ .05), CLES = 64.7\%.
\end{tcolorbox}

A more in-depth analysis could tell how profound these lengthier discussions are, although a higher number of comments might represent developers concerned with the code, willing then to extend their collaboration to the suggestion of refactorings. Previous findings may support those claims; Lee and Cole, when studying the Linux kernel development, acknowledged that the amount of discussion is a quality indicator \cite{Lee_OS_2003}. Also, empirical evidence reports on the impact of the number of comments on changes \cite{Beller_MSR_2014,Pantiuchina_SEM_2020}. 

\subsubsection*{\textbf{H\textsubscript{7}}}
\textbf{Refactoring-inducing and non-refactoring-inducing PRs are equally likely to have a higher number of reviewers} {\small{(AR$_{1}$, AR$_{7}$)}}.

\begin{tcolorbox}[colback = white, left=0pt, right=0pt, top=0pt, bottom=0pt]
\textbf{Finding 7}: We  found  no  statistical  evidence  that  the  number  of  reviewers is  related  to  refactoring-inducement (U = $0.40 \times e^{+06}$, p $< $ .05), CLES = 55.9\%.
\end{tcolorbox}
\vspace{-0.5mm}

Refactoring-inducing and non-refactoring-inducing PRs present two reviewers as median -- the same result found by Rigby et al. \cite{Rigby_Softw_2012} in the OSS scenario. 
There are outliers that, in turn, could be justified by other technical factors, such as complexity of changes, as argued in \cite{Rigby_FSE_2013}. However, our study does not address that scope.

\subsubsection*{\textbf{H\textsubscript{8}}}
\textbf{Refactoring-inducing PRs are more likely to take a longer time to merge than non-refactoring-inducing PRs} {\small{(AR$_{4}$, AR$_{6}$)}}.

\begin{tcolorbox}[colback = white, left=0pt, right=0pt, top=0pt, bottom=0pt]
\textbf{Finding 8}: Refactoring-inducing PRs take significantly more time to merge than non-refactoring-inducing PRs (U = $0.42 \times e^{+06}$, p $<$ .05), CLES = 59.3\%.
\end{tcolorbox}
\vspace{-0.5mm}

We realize the influence of refactorings on time to merge, concluding that time for reviewing and performing refactoring edits both impact the time to merge.
In special, this conclusion is aligned to Szoke et al., who observed a correlation between implementing refactorings and time \cite{Szoke_Springer_2014}, and from Gousios et al., who found that review comments and discussion affect time to merge a PR \cite{Gousios_ICSE_2014}.

\vspace{-4mm}
\subsection{Is Refactoring Induced by Code Reviews?}
To study this research question, we sampled 228 refactoring-inducing PRs, 57 PRs from each of the
\textit{Low}, \textit{Medium}, \textit{High}, and \textit{Very High} categories encompassing one, two to three, four to seven, and eight to 321 refactoring edits, respectively.
By examining 2,096 review comments and 1,207 discussion comments in the sampled PRs, we found 133 (58.3\%) in which at least one refactoring edit was induced by review comments. Such PRs comprise 815 subsequent commits, and 1,891 detected refactorings, 545 of which were induced by review comments.
Finally, we found that \textit{Rename} (35.8\%) (being \textit{readability} a common motivation cited by reviewers) and \textit{Change Type} (30.3\%) operations are the most induced by review in our stratified sample.

\begin{tcolorbox}[colback = white, left=0pt, right=0pt, top=0pt, bottom=0pt]
\textbf{Finding 9}: In a stratified sample of 228 refactoring-inducing PRs, 133 ones (58.3\%) presented at least one refactoring edit induced by code review. \end{tcolorbox}

\vspace{-5mm}
\subsection{Implications}

\textbf{Researchers}: All our findings, except for Finding 7, indicate that refactoring-inducing and non-refactoring-inducing PRs have different characteristics.
Therefore, we recommend that future experiment designs on MCR with PRs to \textit{make a distinction between refactoring-inducing and non-refactoring-inducing PRs, or consider their different characteristics when sampling PRs}.
Researchers can also use our mined data, developed tools, and research methods to investigate code reviewing in pull-based development.

\textbf{Practitioners}: Our findings indicate that there is no statistical difference in the number of reviewers between refactoring-inducing and non-refactoring-inducing PRs (Finding 7). But, all other findings show that refactoring-inducing PRs are associated with more churn (Finding 2), more file changes (Finding 3), more subsequent commits (Finding 4), more review comments (Finding 5), lengthier discussions (Finding 6), and more time to merge (Finding 8) than non-refactoring-inducing PRs. Thus, \textit{we suggest to PR managers to invite more reviewers when a PR becomes refactoring-inducing}, to share the expected increase in review workload, and, perhaps more importantly, to share the knowledge of design changes caused by subsequent refactorings to more team members.

\textbf{Tool builders}: 
In connection to our implication for practitioners, tool builders can \textit{develop bots} \cite{10.1145/2950290.2983989, software_bots_2018} \textit{that recommend reviewers based on some criteria} \cite{10.1145/3377811.3380335} \textit{when a PR becomes refactoring-inducing, to assist the PR managers in inviting additional reviewers}.
Our findings indicate that refactoring-inducing PRs have higher complexity in churn (Finding 2) and file changes (Finding 3).
\textit{Therefore, it is necessary to help the developers distinguish refactoring edits from non-refactoring edits directly in the GitHub or Gerrit review board, where the reviews are actually taking place.}
In the past, researchers implemented refactoring-awareness in the code diff mechanism of IDEs \cite{Ge:2014, Alves:2014, Ge_VLHCC_2017}.
Even though not directly related to our results, we believe that adding refactoring-awareness directly in the GitHub or Gerrit review board -- such as the refactoring-aware commit review Chrome browser extension \cite{chromeExtension} -- would allow reviewers to trace the refactorings performed throughout the commits of a PR, provide prompt feedback, and concentrate efforts on other aspects of the changes, such as collateral effects of refactorings and proposing specific tests.
This recommendation is in agreement with Gousios et al. \cite{Gousios_ICSE_2016}, who emphasized the need for untangling code changes and supporting change impact analysis directly in the PR interface.

\vspace{-3mm}
\section{Threats to Validity} \label{threats-to-validity}
We elaborated our study design after conducting two case studies to better understand GitHub's PRs and procedures of data mining and refactoring detection.
We carefully defined workflows for our research design procedures to explain all decisions taken, and we systematically structured all procedures aiming at replicability. We performed a rigorous selection of the ARL algorithm and input parameters.
To mitigate researcher bias, our qualitative analysis was performed by three analysts. Despite our efforts to perform an initial calibration, there may be limitations concerning conclusions, since we carried out apart analyses.

Nevertheless the establishment of a chain of evidence for the data interpretation and description of taken decisions in the study design, we did not validate the detected refactorings before data analysis, so expressing a potential threat to construct validity (RQ$_1$ and RQ$_2$).
To overcome this issue, we selected RefactoringMiner, a state-of-the-art refactoring detection tool \cite{Tsantalis_ICSE_2018}.
When addressing RQ$_3$, we validated all detected refactorings in our stratified sample.

Aiming to mitigate the risk related to rebasing constraints in our sample, we excluded the PRs merged with the \textit{rebase and merge} option and the PRs including intermediate \textit{merge commits}. Even so, there are still threats due to other non-previously identified manners to search for rebasing operations.

Furthermore, as already admitted in the refactoring-inducing PR definition, we cannot claim that all refactoring edits were caused by reviewing.
To deal with such limitation, we carried out a qualitative analysis of review comments from 228 randomly selected refactoring-inducing PRs, considering a sample size meeting a confidence level of 95\% and a margin of error of 5\%.
Thus, this empirical study provides a particular motivation for further qualitative investigation of review comments to acquire in-depth knowledge on the influence of reviewing on refactoring-inducing PRs.

It is not suitable to generalize the conclusions, except when considering other OSS projects that follow a geographically distributed development and are aligned to ``the Apache way'' principles \cite{Apache_Way}.
Thus, our findings are exclusively extended to cases that have common characteristics with Apache's projects.

\vspace{-2mm}
\section{Related Work} \label{related-work}
By exploring the motivations and challenges of MCR, Bacchelli and Bird identified the code improvements as one of the objectives of reviewing \cite{Bacchelli_ICSE_2013}. 
A finding confirmed by subsequent study on convergent practices of code review by Rigby and Bird \cite{Rigby_FSE_2013}, Beller et al. \cite{Beller_MSR_2014}, and MacLeod et al. \cite{MacLeod_SW_2018}.
Those findings support us in exploring refactorings as a relevant contribution from code reviewing.

The analysis of the technical aspects of code reviewing has been the focus of several empirical studies, in which a few measures have been considered: the number of reviewers by Jiang et al. \cite{Jiang_MSR_2013}, the review comments by Rigby and Bird \cite{Rigby_FSE_2013} and by Beller et al. \cite{Beller_MSR_2014}, the time to merge by Izquierdo-Cortazar \cite{Izquierdo_MSR_2016}, and the size of change by Baysal et al. \cite{Baysal_ESE_2016}.
They provided the first insights on code reviewing aspects investigated in our study.
Also, studies explored the factors influencing code review quality. Bosu et al. discovered that changes' properties affect the review comments usefulness \cite{Bosu_MSR_2015}.  
Nevertheless, Kononenko et al. carried out an analysis concerning how developers perceive code review quality \cite{Kononenko_ICSE_2016}, and figured out that the thoroughness of feedback is the main influencing factor to code review quality. 
Those results corroborate with the findings on the technical aspects empirically studied in \cite{Jiang_MSR_2013,Rigby_FSE_2013,Beller_MSR_2014,Baysal_ESE_2016}, thus constituting an enriched set of technical aspects for investigation.

Paix\~ao et al. found that refactorings' motivations may emerge from code review and influence the composition of edits and number of reviews by analyzing Gerrit reviews \cite{Paixao_MSR_2020}. 
These findings inspired us towards expanding the knowledge regarding code review aspects in GitHub refactoring-inducing PRs.
Pantiuchina et al. analyzed discussion and commits of merged PRs, containing at least one refactoring in one of their commits, and found that most refactorings are triggered from either the original intents of PRs or discussion \cite{Pantiuchina_SEM_2020}. Those findings are motivating since they indicate the influence that review, at the PR level, has on refactorings.
Our study differs from those previous ones because we distinguished refactoring-inducing PRs from non-refactoring-inducing PRs by exploring both reviewing-related aspects and refactoring-inducement.

\vspace{-2mm}
\section{Concluding Remarks} \label{concluding-remark}
We investigated technical aspects characterizing refactoring-inducing PRs based on data mined from GitHub and refactorings detected by RefactoringMiner.
Our results reveal significant differences between refactoring-inducing and non-refactoring-inducing PRs, and a substantial number refactoring edits induced by code reviewing.
As future work, we suggest (i) a further investigation of review comments aiming to identify patterns/practices that could indicate the refactoring-inducement as a contribution of the code review process to the code submitted within PRs; and (ii) exploration of human aspects of reviewers, aiming to enhance the understanding of refactoring-inducement at the PR level. Replications also are highly welcome, since they can support the elaboration of a theory on refactoring-inducing PRs.

\begin{acks}
    We thank the anonymous reviewers for their suggestions to improve this manuscript; and Hugo Addobbati and Ramon Fragoso for their valuable contributions to the qualitative data analysis. 
    This research was partially supported by the National Council for Scientific and Technological Development (CNPq)/Brazil (process 429250/2018-5). 
\end{acks}

\bibliographystyle{ACM-Reference-Format}
\bibliography{references}

\end{document}